\documentstyle[12pt]{article}
\begin{document}\thispagestyle{empty}\begin{flushright}\noindent
29 October 1996                      \hfill
CECM--96--067\\
OUT--4102--63\\
hep-th/9611004                       \end{flushright}\begin{center}{\Large\bf
Evaluations of $k$-fold Euler/Zagier sums:         \\[3pt]
a compendium of results for arbitrary $k$          }\\\vfill{\large
J.~M.~Borwein                                      \hglue 5mm{\tt
jborwein@cecm.sfu.ca                               }\\\vglue 2mm
D.~M.~Bradley                                      \hglue 5mm{\tt
dbradley@cecm.sfu.ca                               }\\\vglue 3mm
CECM, Simon Fraser University,
Burnaby, B.C. V5A 1S6, Canada                      \\\vglue 2mm{\tt
http://www.cecm.sfu.ca/                            }\\\vglue 5mm
D.~J.~Broadhurst                                   \hglue 5mm{\tt
D.Broadhurst@open.ac.uk                            }\\\vglue 3mm
Physics Department, Open University,
Milton Keynes MK7 6AA, UK                          \\\vglue 2mm{\tt
http://yan.open.ac.uk/                             }}\end{center}\vfill
\noindent
Euler sums (also called Zagier sums) occur within the context of knot theory
and quantum field theory. There are various conjectures related to these sums
whose incompletion is a sign that both the mathematics and physics
communities do not yet completely understand the field. Here, we assemble
results for Euler/Zagier sums (also known as multidimensional zeta/harmonic
sums) of arbitrary depth, including sign alternations. Many of our results
were obtained empirically and are apparently new.  By carefully compiling and
examining a huge data base of high precision numerical evaluations, we can
claim with some confidence that certain classes of results are exhaustive.
While many proofs are lacking, we have sketched derivations of all results
that have so far been proved.                    \begin{center}{\footnotesize
(to appear in {\sl Electronic J. Combinatorics})}\end{center}
\newpage
\newcommand{\df}[2]{\mbox{$\frac{#1}{#2}$}}
\newcommand{\eu}{\stackrel{?}{=}}
\newcommand{\us}{\{1\}}
\newcommand{\ou}{\overline1}
\section{Introduction}

We consider $k$-fold Euler sums~\cite{LE,BBG,BG} (also called Zagier sums)
of arbitrary depth $k$.  These sums occur in a natural way within the
context of knot theory and quantum field theory (see~\cite{DJB} for an
extended bibliography), carrying on a rich tradition of algebra and number
theory as pioneered by Euler.  There are various conjectures related to
these sums (see e.g.\ (\ref{z31}) below) whose incompletion is a sign that
both the mathematics and physics communities do not yet completely
understand the field, whence new results are welcome.

As in~\cite{DJB} we allow for all
possible alternations of signs, with $\sigma_j=\pm1$ in
\begin{equation}
\zeta(s_1,\ldots,s_k;\sigma_1,\ldots,\sigma_k)=\sum_{n_j>n_{j+1}>0}
\quad\prod_{j=1}^{k}\frac{\sigma_j^{n_j}}{n_j^{s_j}}\,,\label{form}
\end{equation}
since alternating Euler sums are essential~\cite{BGK} to the
connection~\cite{DK} of knot theory with quantum field
theory~\cite{BK,BDK}.
The integral representation
\begin{eqnarray}
\zeta(s_1,\ldots,s_k;\sigma_1,\ldots,\sigma_k)
&=&\prod_{j=1}^k
\frac{1}{\Gamma(s_j)} \int_1^\infty \frac{dy_j}{y_j}\,
\frac{(\ln y_j)^{s_j-1}}{\prod_{i=1}^j \sigma_i y_i-1}\,,\label{ir}\\
&=&\prod_{j=1}^k\frac{1}{\Gamma(s_j)} \int_0^\infty
\frac{u_j^{s_j-1}\,du_j}{\tau_j \exp\big(\sum_{i=1}^j u_i\big) - 1}
\end{eqnarray}
generalizes that given in~\cite{REC} for non-alternating sums.  Here,
\begin{equation}
\tau_j := \prod_{i=1}^j\sigma_i.\label{taudef}
\end{equation}

For positive integers $s_j$, each $(\ln y_j)^{s_j-1}/\Gamma(s_j)$
in the integrand of~(\ref{ir})
can be written as an iterated integral of the product
$x_1^{-1}dx_1\cdots x_{s_j}^{-1}dx_{s_j}.$
Thus, we have the alternative $(s_1+s_2+\cdots+s_k)$-dimensional
iterated-integral representation
\begin{equation}
\zeta(s_1,\ldots,s_k;\sigma_1,\ldots,\sigma_k)
=\int_0^1 \Omega^{s_1-1} \omega_1 \Omega^{s_2-1} \omega_2\cdots
\Omega^{s_k-1} \omega_k,
\qquad s_1>1,\label{casir}
\end{equation}
in which the integrand denotes a string of distinct differential 1-forms of
type $\Omega = dx/x$ and
$\omega_j$ is given by
\begin{equation}
\omega_j := \frac{\tau_j\,dx_j}{1-x_j \tau_j}.\label{omegadef}
\end{equation}
Note that~(\ref{casir}) shows that Euler sums form a ring, with a product
of sums given by ternary reshuffles of the 1-forms $dx/x$, $dx/(1-x)$, and
$dx/(1+x)$, just as products of non-alternating sums involve
binary~\cite{CK,ZAG1} reshuffles of $dx/x$ and $dx/(1-x)$.

We shall combine the strings of exponents and signs into a single string,
with $s_j$ in the $j$th position when $\sigma_j=+1$, and $\overline{s}_j$
in the $j$th position when $\sigma_j=-1$.  We denote $n$ repetitions of a
substring by $\{\ldots\}_n$.  Finally, we are obliged to point out that the
notation~(\ref{form}) is not completely standard.  In~\cite{REC}, for
example, the argument list is reversed.  Unfortunately, both notations have
proliferated.

For non-alternating sums, several results are known, notably
the duality relation~\cite{CK}:
\begin{equation}
\zeta(m_1+2,\us_{n_1},\ldots,m_p+2,\us_{n_p})=
\zeta(n_p+2,\us_{m_p},\ldots,n_1+2,\us_{m_1})\,,\label{dual}
\end{equation}
an explicit evaluation\footnote{We mark with $\eu$ conjectures for which we
have overwhelming evidence, but no proof.  For unmarked equalities, we
either cite proofs from the literature, or provide a proof sketch in the
appendix.} of the self-dual case with $m_j=n_j=1$, by Zagier~\cite{ZAG1,ZAG2},
(also cited in~\cite{REC}):
\begin{equation}
\zeta(\{3,1\}_n)\,\eu\,\frac{2\cdot\pi^{4n}}{(4n+2)!}\,,\label{z31}
\end{equation}
and the sum rule~\cite{AG}:
\begin{equation}
\sum_{\stackrel{\scriptstyle n_j>\delta_{j,1}}
{N=\Sigma_j n_j}}\zeta(n_1,n_2,\ldots,n_k)=\zeta(N)\,.\label{sr}
\end{equation}
These, and other results have been recast in the language of graded
commutative rings~\cite{MEH}.

We find that~(\ref{z31}) is the first member of a class of
arbitrary-depth results for self-dual non-alternating sums that evaluate
to rational multiples of powers of $\pi^2$, and that
alternating Euler sums of arbitrary depth have a
comparably rich structure.

\section{Generating functions and relations}

We derived the generating function
\begin{equation}
\sum_{m,n\geq0}x^{m+1}y^{n+1}\zeta(m+2,\us_n)=1-\exp\bigg\{\sum_{k\geq2}
\frac{x^k+y^k-(x+y)^k}{k}\,\zeta(k)\bigg\}\,,\label{m2n}
\end{equation}
for the non-alternating sums in the $p=1$ case of~(\ref{dual}),
and the generators
\begin{equation}
\sum_{n\geq0}x^{s n}\zeta(\{s\}_n)=
\prod_{j\geq1}\bigg(1+\frac{x^s}{j^s}\bigg)=
\exp\bigg\{\sum_{k\geq1}\frac{(-1)^{k-1}x^{s k}\zeta(s k)}{k}\bigg\}\,,
\label{all}
\end{equation}
\begin{eqnarray}
\sum_{n\geq0}x^{s n}\zeta(\{\overline{s}\}_n) &=&
\prod_{j\geq1}\bigg(1+(-1)^j\frac{x^s}{j^s}\bigg)\nonumber\\
&=&\exp\bigg\{ \sum_{k\geq1}
\bigg(\frac{2(x/2)^{2s k-s}\zeta(2s k-s)}{2k-1}
-\frac{x^{s k}\zeta(s k)}{k}\bigg)\bigg\}\,,\label{nall}
\end{eqnarray}
with $\Re(s)>1$ in~(\ref{all}), $\Re(s)>0$ in~(\ref{nall}),
and $\zeta(\{\ldots\}_0)=1$. At $s=1$, generator~(\ref{nall}) becomes
\begin{equation}
A(x)\equiv\sum_{n\geq0}x^n\zeta(\{\ou\}_n)
=\frac{2}{B(1+\frac12x,\frac12-\frac12x)}\,.\label{ax}
\end{equation}

We find, empirically, that cases with alternate alternations of
sign are generated by
\begin{equation}
M(x)\equiv\sum_{n\geq0}\left\{x^{2n}\zeta(\{\ou,1\}_n)
+x^{2n+1}\zeta(\{\ou,1\}_n,\ou)\right\}\,\eu\,
\left|A\left(\df{x}{1+i}\right)\right|^2\,,\label{mx}
\end{equation}
for real $x$. This, in turn, generates~(\ref{z31}), via the convolution
\begin{equation}
\sum_{n\geq0}x^{4n}\zeta(\{3,1\}_n)\,\eu\,M(x)M(-x)\,.\label{MM}
\end{equation}
With a further alternating summation, the result analogous to~(\ref{mx}) is
\begin{eqnarray}
T(x)&\equiv&1+\sum_{n\geq0}\left\{x^{2n+1}\zeta(\ou,\{\ou,1\}_n)
+x^{2n+2}\zeta(\ou,\{\ou,1\}_n,\ou)\right\}\nonumber\\
&\eu&M(x)\left\{1-x\,\Im\,\psi\left(1+\df12\,\df{x}{1+i}\right)-x\,\Im\,\psi
\left(\df12-\df12\,\df{x}{1+i}\right)\right\}\,.\label{tx}\end{eqnarray}
Convolution of~(\ref{tx}), in the manner of~(\ref{MM}),
also generates self-dual non-alternating sums:
\begin{equation}
\sum_{n\geq0}x^{4n+2}\zeta(2,\{1,3\}_n)\,\eu\,1-T(x)T(-x)\,.\label{TT}
\end{equation}

Moreover, we discovered the remarkable two-parameter self-dual result
\begin{equation}
\zeta(\{2\}_m,\{3,\{2\}_m,1,\{2\}_m\}_n)\,\eu\,
\frac{2(m+1)\cdot\pi^{4(m+1)n+2m}}{\left(2\{m+1\}\{2n+1\}\right)!}
\,,\label{tower}
\end{equation}
of which the previously known~\cite{REC} example~(\ref{z31}) is
the $m=0$ case.  David Bailey (personal communication) has
confirmed~(\ref{tower}) for $1\le m, n\le 4$ to $800$ decimal places.

Results for sums with unit exponents are generated by
\begin{eqnarray}
L(x)\equiv\sum_{n\geq0}x^n\zeta(\ou,\us_n)&=&\frac{2^{-x}-1}{x}\,,\label{mn}\\
\sum_{n\geq0}x^n\zeta(\ou,\ou,\us_n)&=&
\sum_{k\geq1}\frac{2^{-k}}{k(x-k)}\,,\label{mmn}\\
\sum_{n\geq0}x^n\zeta(\ou,\us_n,\ou)&\eu&
\sum_{k\geq1}\frac{L(k+x)}{k}\,,\label{mnm}\\
\sum_{m,n\geq0}x^{m+1}y^{n+1}\zeta(\ou,\us_m\ou,\ou,\us_n)&\eu&
\sum_{k\geq1}\bigg\{L(k+x)-L(k)\nonumber\\&&{}
-\frac{L(k+x-y)-L(k-y)}{2^y}\bigg\}\,.\label{mas}
\end{eqnarray}

We also discovered the following reductions to non-alternating sums
and unit-exponent alternating sums:
\begin{eqnarray}
\zeta(\{\overline2,1\}_n)&\eu&8^{-n}\zeta(\{2,1\}_n)
=8^{-n}\zeta(\{3\}_n)\,,\label{m21}\\
\zeta(\ou,\us_m,2,{\us_n})&\eu&
\zeta(\ou,\us_n,\ou,\ou,\us_m)
-\zeta(\ou,\us_{m+n+2})\,,\label{dmas}\\
\zeta(\ou,\ou,\us_m,2,{\us_n})&\eu&
\zeta(\ou,\ou,\us_n,\ou,\ou,\us_m)
-\zeta(\ou,\ou,\us_{m+n+2})\nonumber\\&&{}
+\zeta(\ou,\ou,\us_m)\,\zeta(n+2)\,,\label{amas}\\
\zeta(\ou,\us_m,2,2,\us_n)&\eu&
\zeta(\ou,\us_n,\ou,\ou,\ou,\ou,\us_m)
+\zeta(\ou,\us_{m+n+4})\nonumber\\&&{}
-\zeta(\ou,\us_{n+2},\ou,\ou,{\us_m})
-\zeta(\ou,\us_n,\ou,\ou,\us_{m+2})\,,\label{dmd}\\
\zeta(\ou,\ou,\us_m,2,2,\us_n)&\eu&
\zeta(\ou,\ou,\us_n,\ou,\ou,\ou,\ou,\us_m)
+\zeta(\ou,\ou,\us_{m+n+4})\nonumber\\&&{}
-\zeta(\ou,\ou,\us_{n+2},\ou,\ou,{\us_m})
-\zeta(\ou,\ou,\us_n,\ou,\ou,\us_{m+2})\nonumber\\&&{}
+\zeta(\ou,\ou,\us_m,2)\,\zeta(n+2)\nonumber\\&&{}
-\zeta(\ou,\ou,\us_m)\left\{\zeta(n+4)+\zeta(2,n+2)\right\}\,,\label{dmmd}
\end{eqnarray}
\begin{eqnarray}
\zeta(\overline{m+1},\us_n)&\eu&(-1)^m\sum_{k\leq2^m}\varepsilon_k\,
\zeta(\ou,\us_n,S_k)\,,\label{dm}\\
\zeta(\ou,\overline{m+1},\us_n)&\eu&(-1)^m\sum_{k\leq2^m}\varepsilon_k\,
\zeta(\ou,\ou,\us_n,S_k)\nonumber\\&&{}
-\sum_{p\leq m}(-1)^p\zeta(m-p+2,\us_n)\,\zeta(\overline{p})\,,\label{dmm}
\end{eqnarray}
where the last two involve summation over all $2^m$ unit-exponent substrings
of length $m$, with $\sigma_{k,j}$ as the $j$th sign of substring $S_k$,
and $\varepsilon_k=\prod_{m/2>i\geq0}\,\sigma_{k,m-2i}$,
whose effect is to restrict the innermost $m$
summation variables to alternately odd and even integers.

We remark that~(\ref{all}) reduces~(\ref{m21}) to zetas, and
that~(\ref{mn},\ref{mas}) reduce~(\ref{dmas}) to zetas and
the polylogarithms ${\rm Li}_n(1/2)$.
The $m=1$ case of~(\ref{dm}) is reduced to polylogarithms
by~(\ref{mn},\ref{mnm}). The product terms in~(\ref{amas}) and~(\ref{dmm})
are reduced by~(\ref{mmn}) and~(\ref{m2n}); those in~(\ref{dmmd})
involve terms given by~(\ref{mmn},\ref{amas}).
The analysis of~\cite{DJB} shows that new irreducibles, beyond the
polylogarithms from~(\ref{mn}--\ref{mas}), result from
unit-exponent terms generated by~(\ref{amas},\ref{dmd},\ref{dmmd}),
by~(\ref{dm}) when $m\geq2$, and by~(\ref{dmm}) when $m\geq1$.

\section{Evaluations at arbitrary depth}

{From} the symmetric generator~(\ref{m2n}), we obtain
\begin{eqnarray}
\zeta(2,\us_n)&=&\zeta(n+2)\,,\label{21}\\
\zeta(3,\us_n)&=&\zeta(n+2,1)=\frac{n+2}{2}\,\zeta(n+3)
-\frac12\sum_{k=1}^n\zeta(k+1)\,\zeta(n+2-k)\,,\label{31}
\end{eqnarray}
and, in general, products of up to $\min(m+1,n+1)$ zetas in
$\zeta(m+2,\us_n)=\zeta(n+2,\us_m)$, whose symmetry was known
from~(\ref{dual}). Note that~(\ref{21}) is also implied by~(\ref{sr}).

For integer values, $s=m$, generators~(\ref{all},\ref{nall}) give
\begin{eqnarray}
\sum_{n\geq0}x^{m n}\zeta(\{m\}_n)&=&
\prod_{j=1}^m\frac{1}{\Gamma(1-\omega_m^{2j-1} x)}\,,\label{int}\\
\sum_{n\geq0}x^{m n}\zeta(\{\overline{m}\}_n)&=&
\prod_{j=1}^m\frac{\sqrt{\pi}}{\Gamma(1-\frac12\omega_m^{2j-1}x)\,\Gamma
(\frac12-\frac12\omega_m^{2j}x)}\,,\label{nint}
\end{eqnarray}
\noindent with $\omega_m=\exp(i\pi/m)$.

For even integers, $m=2p$, generators~(\ref{int},\ref{nint})
give trigonometric products:
\begin{eqnarray}
S_p(x)\equiv\sum_{n\geq0}x^{2p n}\zeta(\{2p\}_n)&=&
(i\pi x)^{-p}\prod_{j=1}^p\sin(\pi\omega_{2p}^{2j-1}x)\,,\label{sin}\\
\sum_{n\geq0}x^{2p n}\zeta(\{\overline{2p}\}_n)&=&S_p(\df12x)\prod_{j=1}^p
\cos(\df12\pi\omega_p^j x)\,,\label{cos}
\end{eqnarray}
which show that
$\zeta(\{2p\}_n)$ and $\zeta(\{\overline{2p}\}_n)$
are rational multiples of $\pi^{2p n}$.

The non-alternating result~(\ref{sin}) readily yields
\begin{eqnarray}
\zeta(\{2\}_n)&=&\frac{2\cdot(2\pi)^{2n}}{(2n+1)!}
\bigg(\frac12\bigg)^{2n+1}\,,\label{r2}\\
\zeta(\{4\}_n)&=&\frac{4\cdot(2\pi)^{4n}}{(4n+2)!}
\bigg(\frac12\bigg)^{2n+1}\,,\label{r4}\\
\zeta(\{6\}_n)&=&\frac{6\cdot(2\pi)^{6n}}{(6n+3)!}\,,\label{r6}\\
\zeta(\{8\}_n)&=&\frac{8\cdot(2\pi)^{8n}}{(8n+4)!}
\bigg\{\bigg(1+\frac{1}{\sqrt2}\bigg)^{4n+2}
+\bigg(1-\frac{1}{\sqrt2}\bigg)^{4n+2}\bigg\}\,.\label{r8}
\end{eqnarray}
Comparison of~(\ref{r4}) with~(\ref{z31}) reveals that
Zagier's conjecture can
be reformulated as
\begin{equation}
4^n\zeta(\{3,1\}_n) \eu \zeta(\{4\}_n)
\end{equation}
or, in the notation of~(\ref{casir}),
\begin{equation}
4^n \int_0^1(\Omega^2\omega^2)^n \eu \int_0^1(\Omega^3\omega)^n.\label{Zcon1}
\end{equation}
Equivalently, from~(\ref{r2}), it becomes
\begin{equation}
(2n+1)\zeta(\{3,1\}_n) \eu \zeta(\{2,2\}_n)
\end{equation}
or
\begin{equation}
(2n+1)\int_0^1(\Omega^2\omega^2)^n \eu
\int_0^1(\Omega\omega)^{2n},\label{Zcon2}
\end{equation}
in which, unlike~(\ref{Zcon1}), the list of omegas is merely reordered.
Comparison of the empirical result~(\ref{tower})
with~(\ref{r2},\ref{r4}) reveals that
\begin{eqnarray}
\zeta(\{2\}_m,\{3,\{2\}_m,1,\{2\}_m\}_n)&\eu&
\frac{1}{2n+1}\,\zeta(\{2\}_{2(m+1)n+m})
\,,\label{tower2}\\
\zeta(\{2\}_{2p},\{3,\{2\}_{2p},1,\{2\}_{2p}\}_n)&\eu&
\frac{2p+1}{4^{(2p+1)n+p}}\,\zeta(\{4\}_{(2p+1)n+p})\,.\label{tower4}
\end{eqnarray}
Result~(\ref{r8}) was already known~\cite{PC}.
The next member of the series is rather beautiful:
\begin{equation}
\zeta(\{10\}_n)=\frac{10\cdot(2\pi)^{10n}(L_{10n+5}+1)}{(10n+5)!}\,,
\label{r10}
\end{equation}
where $L_n=L_{n-1}+L_{n-2}$ is the $n$th Lucas number,
with $L_1=1$ and $L_2=3$.

In the general case, a Laplace transform of~(\ref{sin}) yields
\begin{equation}
\sum_{n\geq0}(2p n+p)!\left(\frac{z}{(2\pi)^p}\right)^n\zeta(\{2p\}_n)
=2p\sum_{k=1}^{N_{p}}\frac{z_{p,k}^{1/2}}{z_{p,k}-z}\,,\label{ib}
\end{equation}
\noindent with $N_p\leq2^p/2p$ poles, whose positions
$\{z_{p,k}\mid 1\leq k\leq N_p\}$
are determined by the Laplace transforms of the $2^p$ exponentials generated
by the product in~(\ref{sin}). The pole closest to the origin, at $z_{p,1}=
(2\sin(\pi/2p))^{2p}$, gives the first term in
\begin{equation}
\zeta(\{2p\}_n)=\frac{2p\cdot(2\pi)^{2p n}}{(2p n+p)!}
\left(\frac{1}{2\sin\frac{\pi}{2p}}\right)^{2p n+p}\bigg\{1+
\sum_{k=2}^{N_{p}}R_{p,k}^{2p n+p}\bigg\}\,,\label{app}
\end{equation}
with $R_{p,k}=(z_{p,1}/z_{p,k})^{1/2p}$, and hence $|\,R_{p,k}|<1$ for $k>1$.
Choices of signs, $\sigma_j=\pm1$, in
\begin{equation}
\frac{|\,R_{p,k}|}{\sin\frac{\pi}{2p}}=
\bigg|\sum_{j=1}^p\sigma_j\omega_p^j\,\bigg|\,,\label{set}
\end{equation}
yield all the absolute values, though some choices of sign may not be
realized in~(\ref{app}).

Proceeding up to $p=9$, we derived:
\begin{eqnarray}
\zeta(\{12\}_n)&=&\frac{12\cdot(2\pi)^{12n}}{(12n+6)!}
\bigg\{\bigg(\frac{1+\sqrt3}{\sqrt2}\bigg)^{12n+6}
+\bigg(\frac{1-\sqrt3}{\sqrt2}\bigg)^{12n+6}+2^{6n+3}\bigg\}\,,
\label{r12}\\
\zeta(\{14\}_n)&=&\frac{14\cdot(2\pi)^{14n}}{(14n+7)!}\,\Re\bigg(
\sum_{k=1}^3\frac{1+r_k^{28n+14}}{r_k^{14n+7}}
+2\bigg(\frac{i\sqrt7-1}{2}\bigg)^{14n+7}+1\bigg)\,,\label{r14}\\
\zeta(\{16\}_n)&=&\frac{16\cdot(2\pi)^{16n}}{(16n+8)!}
\sum_{k=1}^4\Re\left(\frac{1}{s_k^{16n+8}}+\frac{s_k^{16n+8}}{c_k^{16n+8}}
+2\left(\frac{i}{c_k}+c_k+\sqrt2\right)^{8n+4}\right)\,,\label{r16}\\
\zeta(\{18\}_n)&=&\frac{18\cdot(2\pi)^{18n}}{(18n+9)!}
\sum_{k=1}^3\Re\left(\frac{1}{t_k^{18n+9}}+(1+t_k)^{18n+9}
+2\left(-\omega_3-t_k\right)^{18n+9}\right)\,.\label{r18}
\end{eqnarray}
In~(\ref{r14}), $r_k=2\cos((2k-1)\pi/7)$ are the roots of
the cubic equation $r(1+r)(2-r)=1$.
In~(\ref{r16}), $s_k=2\sin((2k-1)\pi/16)$ and
$c_k=2-s_k^2$, which are the roots of $(2-c^2)^2=2$.
In~(\ref{r18}), $t_k=2\cos(2^k\pi/9)$ are the roots of $t(3-t^2)=1$.
The method adopted to obtain these results
exploited the exactness of the $[N-1\backslash N]$ Pad\'e approximant
to~(\ref{ib}), for $N\geq N_{p}$. The roots of its denominator
were then used to find $R_{p,k}=2\sin(\pi/2p)/z_{p,k}^{1/2p}$.

The $p$-th member of the integer sequence\footnote{The integer
sequence~(\ref{pos}) was not identified by Neil Sloane's `superseeker'
utility~\cite{NJAS}.}
\begin{equation}
1,1,1,2,3,4,8,12,16,
33,62,67,186,316,280,1040,1963,1702,6830,10751,\ldots\label{pos}
\end{equation}
\noindent gives the number of distinct non-zero absolute values of
$\sum_{j=1}^p\sigma_j\omega_p^j$.
Of these possibilities,
\begin{equation}
1,1,1,2,3,3,8,12,9,\ldots\label{get}
\end{equation}
are present in~(\ref{app}). Hence, for $p=6$ and $p=9$,
some of the choices of signs in~(\ref{set}) are absent.
Correspondingly, the values of $N_{p}$ in the sequence
\begin{equation}
1,1,1,2,3,3,9,16,12,\ldots\label{Np}
\end{equation}
do not saturate the upper bound $\lfloor2^p/2p\rfloor$, for $p=6$
and $p=9$.

Explicit results from~(\ref{cos}) are much lengthier
than those from~(\ref{sin}), since the former gives $4^p$ exponentials,
while the latter gives only $2^p$. We cite only the first three cases:
\begin{eqnarray}
\zeta(\{\overline{2}\}_n)&=&\frac{\pi^{2n}}{(2n+1)!}
\,\frac{(-1)^{n(n+1)/2}}{2^n}\,,\label{m2}\\
\zeta(\{\overline{4}\}_n)&=&\frac{\pi^{4n}}{(4n+2)!}
\,\frac{(-1)^{n(n+1)/2}}{2^n}\,\bigg((1+\sqrt2)^{2n+1}+(1-\sqrt2)^{2n+1}
\bigg)\,,\label{m4}\\
\zeta(\{\overline{6}\}_n)&=&\frac{\pi^{6n}}{(6n+3)!}\cdot\frac32
\bigg(1+2^{3n+1}(-1)^{n(n+1)/2}\\
&\qquad&\qquad\times\bigg\{\bigg(\frac{1+\sqrt3}{2}\bigg)^{6n+3}+
\bigg(\frac{1-\sqrt3}{2}\bigg)^{6n+3}-1\bigg\}\bigg)\,.\label{m6}
\end{eqnarray}
Comparison of~(\ref{r2}) with~(\ref{m2}) reveals that
\begin{equation}
\zeta(\{\overline{2}\}_n)=2^{-n}(-1)^{\lceil n/2\rceil}
\zeta(\{{2}\}_n)\,.\label{ceil}
\end{equation}
Finally, from~(\ref{nall}) we obtain
\begin{equation}
\zeta(\{\ou\}_n) = (-1)^n \sum \prod_{k\ge 1}
\frac{1}{j_k!}\bigg(\frac{-{\rm Li}_k((-1)^k)}{k}\bigg)^{j_k},
\end{equation}
where the sum is over all non-negative integers satisfying $\sum_{k\ge 1} k
j_k = n$.

{From}~(\ref{TT}), we obtain
a self-dual evaluation, more complex than~(\ref{tower}):
\begin{eqnarray}
\zeta(2,\{1,3\}_n)&\eu&4^{-n}\sum_{k=0}^n(-1)^k\zeta(\{4\}_{n-k})
\bigg\{(4k+1)\,\zeta(4k+2)\nonumber\\&&{}
-4\sum_{j=1}^k\zeta(4j-1)\,\zeta(4k-4j+3)\bigg\}\,,\label{2134}
\end{eqnarray}
with $\pi^2$ terms generated by $\zeta(4k+2)$ and by~(\ref{r4}).
The absence of $\zeta(4k+1)$ is conspicuous.

Explicit results generated by~(\ref{mn}--\ref{mas}) involve the
polylogarithms
\begin{equation}
A_n\equiv{\rm Li}_n(1/2)=\sum_{k=1}^\infty\frac{1}{2^k k^n}\,,\quad
P_n\equiv\frac{(\ln2)^n}{n!}\,,\quad Z_n\equiv(-1)^n\zeta(n)\,,\label{apz}
\end{equation}
\noindent in terms of which we obtain
\begin{eqnarray}
\zeta(\ou,\us_n)&=&(-1)^{n+1}P_{n+1}\,,\label{emn}\\
\zeta(\ou,\ou,\us_n)&=&-A_{n+2}\,,\label{emmn}\\
\zeta(\ou,\us_n,\ou)&\eu&-Z_{n+2}
+(-1)^n\sum_{k=1}^{n+2}A_k P_{n+2-k}\,,\label{emnm}\\
\zeta(\ou,\us_m,\ou,\ou,\us_n)&\eu&
(-1)^m\sum_{k=1}^{m+2}{n+k\choose n+1}A_{k+n+1}P_{m+2-k}\nonumber\\&+&
(-1)^n\sum_{k=1}^{n+2}{m+k\choose m+1}Z_{k+m+1}P_{n+2-k}\,.
\label{emas}
\end{eqnarray}

\section{Evaluations at specific depths}

Several thousand evaluations, obtained in the work for~\cite{DJB} with the
aid of MPPSLQ~\cite{DHB} and REDUCE~\cite{AH}, were inspected, in a search
for further, comparably simple, results. These include analytical results
for all 1457 sums with weight $w=\sum_j s_j\leq7$, for all 3698 double sums
with weight $w\leq44$, and for all 1092 non-alternating sums with depth
$k\leq4$ and weight $w\leq14$. To these we adjoined more than 2000
strategically selected high-precision numerical evaluations of self-dual
sums with $s_j\leq3$ and weights up to $w=40$, which enabled the discovery
and validation of the remarkable generalization of~(\ref{z31}) that is
given in~(\ref{tower}).  The reader will find a detailed discussion of our
scheme for computing these high-precision numerical evaluations in section
4 of~\cite{DJB}.  For other approaches, see~\cite{CranBuhl} and~\cite{RC}
in which Euler-Maclaurin based techniques are eschewed in favour of
transformation to explicitly convergent sums.

It was found that precisely 11 of the 64 convergent depth-7 sums with unit
exponents are reducible to the polylogarithms~(\ref{apz}) and their products.
They are given by the
6 results~(\ref{ax},\ref{mx},\ref{tx},\ref{emn},\ref{emmn},\ref{emnm})
and 5 instances of~(\ref{emas}). Combining these with
5 instances of~(\ref{dmas}) and the $m=1$ case of~(\ref{dm}),
we exhaust the weight-7 reducible alternating sums with depth $k\geq5$.
We computed, to high precision, all 2046 self-dual non-alternating sums
comprising up to 10 `atomic' substrings
of the form $\{m+2,\{1\}_n\}$, with $m,n=0,1$, as in~(\ref{tower},\ref{2134}),
and hence having weight $w=2k\leq40$. Precisely 25 of these
are rational multiples of powers of $\pi^2$. They are exhausted
by~(\ref{tower}). Moreover,~(\ref{m2n},\ref{tower},\ref{2134}) were found
to exhaust all zeta-reducible cases of non-alternating sums with $w=2k=10$,
of self-dual sums with $w=12$, and of self-dual sums with $s_j\leq3$
and $8\leq w\leq16$. At $w=16$, computation and MPPSLQ
analysis of 34 self-dual sums, to 300 significant figures,
took about 0.5 CPUhour/sum on a DEC AlphaStation 600 5/333 at the
Open University. Such exhaustion of reducible cases by our
results~(\ref{m2n}--\ref{dmm}) suggests that they are, like our database,
reasonably comprehensive.

Among many MPPSLQ results at specific depths, the following
are rather distinctive:
\begin{eqnarray}
\zeta(2,1,\overline2,\overline2)&\eu&\df{39}{128}\,\zeta(4)\,\zeta(3)
-\df{193}{64}\,\zeta(5)\,\zeta(2)+\df{593}{128}\,\zeta(7)\,,\label{rm1}\\
\zeta(\overline2,\overline2,1,2)&\eu&\df{9}{128}\,\zeta(4)\,\zeta(3)
+\df{447}{128}\,\zeta(5)\,\zeta(2)-\df{1537}{256}\,\zeta(7)\,,\label{rm2}\\
\zeta(\{4,1,1\}_2)&\eu&\df{3\pi^4}{16}\left\{\zeta(6,2)-4\zeta(5)\,\zeta(3)
\right\} -\df{41\pi^6}{5040}\left\{\zeta^2(3)-\df{77023\pi^6}{14414400}\right
\}\nonumber\\&&{}+\df{397}{8}\zeta(9)\,\zeta(3)+\zeta^4(3)\,,\label{l12a}\\
\zeta(2,2,1,2,3,2)&\eu&\df{75\pi^2}{32}\left\{\zeta(8,2)-2\zeta(7)\,
\zeta(3)+\df{34}{225}\zeta^2(5)+\df{4528801\pi^{10}}{61297236000}\right\}
\nonumber\\&&{}-\df{825}{8}\zeta(7)\,\zeta(5)\,,\label{l12b}\\
\zeta(\{\overline3,1\}_2)&\eu&-7\left(\alpha(5)-\df{39}{64}\zeta(5)+\df18\zeta
(4)\ln2\right)\zeta(3)+\left(2\alpha(4)-\df14\zeta(4)\right)^2\nonumber\\&&{}
+2\left(\alpha(4)-\df{15}{16}\zeta(4)+\df78\zeta(3)\ln2\right)^2
-\df{1}{32}\zeta(8)\,,\label{l8}
\end{eqnarray}
with $\alpha(n)\equiv A_n+(-1)^n(P_n-\frac{\pi^2}{12}\,P_{n-2})$,
as in~\cite{DJB}.
Note that the alternating sums~(\ref{rm1},\ref{rm2}) are pure zeta,
yet we were unable to find generalizations of them; only
from~(\ref{nall},\ref{m21}) have we obtained arbitrary-depth
pure-zeta alternating results.
Note also that the self-dual sums~(\ref{l12a}) and~(\ref{l12b}),
with $w=2k=12$,
contain non-zeta~\cite{BBG} irreducibles, $\zeta(6,2)$ and $\zeta(8,2)$,
yet their kinship with distinct reducible classes, generated
by~(\ref{MM}) and~(\ref{TT}), manifests itself in the unusual circumstance
that they share only $\pi^{12}$ as a common term. Finally, note
that the polylogarithmic complexity of~(\ref{l8})
contrasts greatly with the zeta-reducibility of~(\ref{m21}), via~(\ref{all}),
yet its kinship with~(\ref{m21}) is reflected by the absence of 12 of the
21 terms~\cite{DJB} that occur in alternating sums with $w=2k=8$.
In each of~(\ref{rm1}--\ref{l8}) one senses, from the relatively small
number of terms, a degree of proximity to an arbitrary-depth reduction.

It is conjectured that, at any depth $k>1$, Euler sums of weight $w$ are
reducible to a rational linear combination of lesser-depth sums (and their
products) whenever $w$ and $k$ are of opposite parity. It is also
conjectured that the lowest-weight irreducible depth-$k$ alternating sum
occurs at weight $k+2$ and entails ${\rm Li}_{k+2}(1/2)$~\cite{DJB}. The
critical weight $w_k$, at which depth-$k$ non-alternating sums first fail
to be reducible to non-alternating sums of lesser depth, is more
problematic. In~\cite{BBG} it was found that $w_2=8$; in~\cite{BG} that
$w_3=11$; in~\cite{DJB} that $w_4=12$. Reducibility was proved
below these critical weights; reducibility at them was
shown to be incredible, by lattice methods~\cite{DHB}.
There is likewise good support for $w_5=15$ and $w_6=18$.
It is conjectured~\cite{DJB2} that $w_k=3k$,
for all $k\geq4$. It appears that a large majority
of non-alternating sums are irreducible
whenever $w$ and $k$ are of the same parity and $w\ge w_k$. Additionally,
R. Girgensohn (personal communication) has outlined a proof that,
in the notation of~(\ref{form}),
\[\zeta(s_1,\ldots,s_k;\sigma_1,\ldots,\sigma_k)+(-1)^k
\zeta(s_k,\ldots,s_1;\sigma_k,\ldots,\sigma_1)\] is reducible for every
$k>1$.

For depths 2, 3 and 4, we have the following more specific
remarks:

\noindent
{\bf Depth 2.}
Whenever $s+t$ is odd, we have
\begin{equation}
  \zeta(s,t;\sigma,\tau)  =
  \df{1}{2}\left(-\lambda_{s+t}+(1+(-1)^s)\zeta(s;\sigma)\zeta(t;\tau) +
  \mu_{s+t}\right) - \sum_{0<k<(s+t)/2} \lambda_{2k} \mu_{s+t-2k},
\end{equation}
where $\lambda_r = \zeta(r;\sigma\tau)$ and  $\mu_r = (-1)^s \left({r-1
\choose s-1}\zeta(r;\sigma) + {r-1\choose t-1}\zeta(r;\tau)\right).$ This
compact formula summarizes the evaluations given in~\cite{BG}.  Recently, a
shorter proof has been given by R. Girgensohn (personal communication). A
conjectured minimal {\bf Q}-basis for all depth-2 Euler sums is formed
by~\cite{DJB}:
the depth-1 sums,
$\ln2$, $\pi^2$, $\{\zeta(2a+1)\mid a>0\}$, and the depth-2 sums
$\{\zeta(\overline{2a+1},\overline{2b+1})\mid a>b\ge0\}$.
All 3698 convergent double sums with weights $w\leq44$ have been
proved~\cite{DJB} to
be expressible in this basis, using identities derived in~\cite{BBG}
and augmented in~\cite{DJB}. A conjectured minimal {\bf Q}-basis for
non-alternating depth-2 Euler sums is formed by $\pi^2$, $\{\zeta(2a+1)\mid
a>0\}$ and $\{\zeta(2a+1,2b+1)\mid a\ge 2b>0\}$, which is likewise proven
to be sufficient up to weight 44. It is conjectured that the proven
result~\cite{BBG}
\begin{equation}
   \zeta(4,2) = \zeta^2(3)-\frac{4\pi^6}{2835},
\end{equation}
is the {\em sole\/} case of an even-weight reduction of a
non-alternating sum $\zeta(a,b)$ with $a>b>1$.

\noindent {\bf Depth 3.}
In~\cite{BG}, it is proved that non-alternating Euler sums of depth 3 and
weight $w$ are reducible to a rational linear combination of lesser depth
sums when $w$ is even or $w\le10$.  It is conjectured that most depth-3
non-alternating sums of odd weight exceeding 10 are irreducible. The only
reductions that  have been found at odd weights in the range 17 to 33 are
the cases $\zeta(a,a,a)$ and $\zeta(a,1,1)$. A conjectured {\bf Q}-basis
for all depth-3 non-alternating sums is the set of lesser-depth
non-alternating sums along with the set $\{ \zeta(2a+1,2b+1,2c+1) \mid a\ge
b\ge c>0, a>c\}$.

\noindent {\bf Depth 4.}
It is proved~\cite{DJB2} that every depth-4 non-alternating Euler sum with
weight less than 12 is reducible to non-alternating sums of lesser depth.
It is conjectured that a depth-4 non-alternating Euler sum with even weight
exceeding 14 is reducible if and only if it is of one of the following
forms: $\zeta(a,b,a,1)$, $\zeta(a,a,1,b)$, $\zeta(a,1,b,b)$,
or $\zeta(a,b,b,a)$,
with $a=b$, or $b=1$, permitted. (It is proven and will be shown in a
subsequent paper that these forms reduce.)

For more on questions of reducibility, see~\cite{DJB,DJB2}.

\section{Conclusions}

Euler sums of arbitrary depth are a rich source of fascinating identities,
with~(\ref{tx}) and~(\ref{tower}) serving as spectacular examples.
Many of our results were discovered empirically;
to date, we have not proven
conjectures~(\ref{mx}--\ref{tower}, \ref{mnm}--\ref{dmm})
and their corollaries.
The evidence in their favour is, however, overwhelming.
The reader may consult the appendix for sketched derivations of
results that have been proved.

\noindent{\bf Acknowledgements.}
We thank Dirk Kreimer for informing us
that~(\ref{dual}) is in~\cite{CK},
Richard Crandall for telling us about~(\ref{z31}) and~(\ref{r8}),
and Chris Stoddart for skillful computer management.

\section{Appendix: Some Proof Sketches}

The integral representation~(\ref{ir}) may be derived using the well-known
identity
\begin{equation}
n^{-s}\Gamma(s) = \int_1^\infty (\log y)^{s-1} y^{-n-1}\,dy.
\end{equation}
Thus, the LHS of~(\ref{ir}) may be written as
\begin{equation}
\zeta(s_1,\ldots,s_k;\sigma_1,\ldots,\sigma_k)
= \sum \prod_{j=1}^k \int_1^\infty \frac{d y_j}{y_j}\,
\frac{(\log y_j)^{s_j-1}}{\Gamma(s_j)}\bigg(\frac{\sigma_j}{y_j}\bigg)^{n_j},
\label{deriveir}
\end{equation}
where the sum is over all positive integers $n_1>n_2>\cdots>n_k>0.$
Now make the change of summation variables $m_k=n_k$, and $m_j = n_j-n_{j+1}$
for $j=1,2,\ldots,k-1.$  Then each $m_j$ runs independently over the positive
integers, and~(\ref{deriveir}) becomes
\begin{eqnarray}
\zeta(s_1,\ldots,s_k;\sigma_1,\ldots,\sigma_k)
&=& \prod_{j=1}^k \int_1^\infty \frac{d y_j}{y_j}\,
\frac{(\log y_j)^{s_j-1}}{\Gamma(s_j)}\sum_{m_j\ge 1} \bigg(\prod_{i=1}^j
\frac{\sigma_i}{y_i}\bigg)^{m_j}\nonumber\\
&=& \prod_{j=1}^k\frac{1}{\Gamma(s_j)} \int_1^\infty \frac{d y_j}{y_j}\,
\frac{(\log y_j)^{s_j-1}}{\prod_{i=1}^j y_i/\sigma_i - 1},
\end{eqnarray}
after summing the geometric series.  Since each $\sigma_i=\pm 1$, this
is the same as~(\ref{ir}).

In the introduction, we briefly indicated how the iterated-integral
representation~(\ref{casir}) arises from the non-iterated multiple integral
representation~(\ref{ir}).  We present a direct derivation below.  Yet
another approach is taken in~\cite{CK}, but there only the non-alternating
case is considered. With $\Omega$ and $\omega_j$ as in the introduction,
put $\Omega_n:=x^n\Omega = x^n \,dx/x.$  We begin with the self-evident
integral representation
\begin{equation}
\frac{y^n}{n^k} = \int_0^y \Omega^{k-1} \Omega_n,\label{start}
\end{equation}
valid for positive integers $n$ and $k$.
It follows that for positive integers $n$, $p$, $r$, and $k$,
\begin{equation}
\frac{y^{n+p}}{(n+p)^r n^k} = \frac{1}{n^k}\int_0^y \Omega^{r-1}\Omega_{n+p}
=\int_0^y \Omega^{r-1} \bigg(\frac{x^n}{n^k}\bigg) x^p\frac{dx}{x}.
\end{equation}
Now substitute~(\ref{start}) for $x^n/n^k$, obtaining
\begin{equation}
\frac{y^n}{n^r (n+p)^k}
=\int_0^y \Omega^{r-1} \int_0^x \Omega^{k-1}\Omega_n\,x^p\frac{dx}{x}
=\int_0^y \Omega^{r-1} \Omega_p\Omega^{k-1}\Omega_n.
\end{equation}
In general, for positive integers $m_j$, $s_j$, we have
\begin{equation}
\frac{y^{m_1}}{\prod_{j=1}^k (m_1+m_2+\cdots +m_j)^{s_j}}
= \int_0^y \prod_{j=1}^k\Omega^{s_j-1}\Omega_{m_j}.\label{omegas}
\end{equation}
But, recalling the definition~(\ref{taudef}) of $\tau_j$ from the
introduction, we have
\begin{equation}
\zeta(s_1,\ldots,s_k;\sigma_1,\ldots,\sigma_k)
=\sum_{n_j>n_{j+1}}\prod_{j=1}^k \frac{\sigma_j^{n_j}}{{n_j}^{s_j}}
=\sum_{m_j\ge 1}\prod_{j=1}^k \frac{\tau_j^{m_j}}{(m_1+m_2+\cdots+m_j)^{s_j}}.
\end{equation}
Thus, from~(\ref{omegas}),
\begin{equation}
\zeta(s_1,\ldots,s_k;\sigma_1,\ldots,\sigma_k)
= \sum_{m_j\ge 1} \int_0^1 \prod_{j=1}^k \Omega^{s_j-1}
\tau_j^{m_j}\Omega_{m_j}
= \int_0^1 \prod_{j=1}^k \Omega^{s_j-1}\omega_j,\label{casirdone}
\end{equation}
by summing the $k$ geometric series and recalling the
definition~(\ref{omegadef}) of $\omega_j$ from the introduction.

A general property of iterated integrals~\cite{CK} such as~(\ref{casir})
or~(\ref{casirdone}) is that the string in the integrand can be reversed if
the integration limits are exchanged and the appropriate sign factor is
taken into account.  If in addition, the integration variables $x_j$ are
all replaced by their complement $1-x_j$, this has the effect of switching
$\Omega$ and $\omega$.  Thus,
\begin{eqnarray}
\zeta(m_1+2,\us_{n_1},\ldots,m_p+2,\us_{n_p}) &=& \int_0^1
\Omega^{m_1+1}\omega^{n_1+1}\cdots\Omega^{m_p+1}\omega^{n_p+1}\nonumber\\
 &=& \int_0^1
 \Omega^{n_p+1}\omega^{m_p+1}\cdots\Omega^{m_1+1}\omega^{n_1+1}\nonumber\\
 &=& \zeta(n_p+2,\us_{m_p},\ldots,n_1+2,\us_{m_1}),
\end{eqnarray}
which proves the duality relation~(\ref{dual}).

To prove~(\ref{m2n}), we write the left side as
\begin{equation}
x y\sum_{m\ge 0}\sum_{k\ge
1}\frac{x^m}{k^{m+2}}\prod_{j=1}^{k-1}\bigg(1+\frac{y}{j}\bigg).
\end{equation}
After summing on $m$, what remains is an instance of the hypergeometric series
with first term omitted:
\begin{equation}
1-{}_2F_1(-x,y;1-x) = 1-\frac{\Gamma(1-x)\Gamma(1-y)}{\Gamma(1-x-y)}, \quad
\Re(x+y)<1.
\end{equation}
To complete the proof, write $\Gamma$ in the form
$\exp(\int\Gamma'/\Gamma)$ and employ the Maclaurin series representation
for $\Gamma'/\Gamma$.

For~(\ref{all}), write
\begin{equation}
F(x) := \sum_{n\ge 0}x^{sn}\zeta(\{s\}_n)
\end{equation}
and note that
\begin{equation}
F(x) = \prod_{j\ge 1}\bigg(1+\frac{x^s}{j^s}\bigg)\label{Fprod}
\end{equation}
follows directly from the definition~(\ref{form}).  Taking the logarithmic
derivative, we have
\begin{equation}
\frac{F^\prime}{F}(x) = \sum_{j\ge 1}\frac{s
x^{s-1}/j^s}{(1+x^s/j^s)}.\label{logder}
\end{equation}
Now expand the denominator of~(\ref{logder}) in powers of $x^s/j^s$ and
interchange summation
order, obtaining
\begin{equation}
\frac{F'}{F}(x) = \sum_{k\ge 1}(-1)^{k-1}s x^{s k-1}\zeta(sk).
\end{equation}
Finally, integrate, exponentiate, and check that the result agrees
with~(\ref{Fprod}) at $x=0.$

The proof of~(\ref{nall}) is analogous, with
\begin{equation}
G(x) := \prod_{j\ge 1}\bigg(1+(-1)^j \frac{x^s}{j^s}\bigg)
\end{equation}
replacing $F(x)$ in~(\ref{Fprod}) above.  Note that the special
case~(\ref{ax}) is example 1, page 259 of \cite{WW}.

Although we currently have no proof of~(\ref{mx}), from
\begin{eqnarray}
A\bigg(\frac{x}{1+i}\bigg)A\bigg(\frac{x}{1-i}\bigg)
&=& \prod_{j\ge 1} \bigg(1 + \frac{(-1)^j x}{j\sqrt2}e^{i\pi/4}\bigg)
      \bigg(1+\frac{(-1)^j x}{j\sqrt2} e^{-i\pi/4}\bigg)\\
&=& \prod_{j\ge 1}\bigg(1+\frac{(-1)^j x}{j} + \frac{x^2}{2j^2}\bigg),
\end{eqnarray}
it follows that
\begin{equation}
\bigg(\frac{1}{2i}\bigg)^n\sum_{k=0}^{2n} i^k
\zeta(\{\ou\}_k)\zeta(\{\ou\}_{2n-k}) = \sum_{2p + q=2n}\zeta(\{\ou\}_q)
2^{-p}\zeta(\{2\}_p).
\end{equation}
Similarly, from
\begin{equation}
\prod_{j\ge 1}\bigg(1+\frac{(-1)^j x^3}{j^3}\bigg)  = \prod_{j\ge
1}\bigg(1+\frac{(-1)^j x}{j}\bigg)\bigg(1-\frac{(-1)^j
x}{j}+\frac{x^2}{j^2}\bigg),
\end{equation}
it follows that
\begin{equation}
\zeta(\{\overline3\}_n) = \sum_{2p + q+r=3n}
(-1)^q\zeta(\{\ou\}_q)\zeta(\{\ou\}_r)\zeta(\{2\}_p).
\end{equation}

To prove~(\ref{mn}), take $t=-1$ in
\begin{equation}
\sum_{m\ge 1}\frac{t^m}{m}\prod_{j=1}^{m-1}\bigg(1+\frac{x}{j}\bigg)
= t\sum_{m\ge 0} (-t)^m {-x-1\choose m}\int_0^1 u^m \,du
= \frac{(1-t)^{-x}-1}{x}.\label{pfmn}
\end{equation}

For~(\ref{mmn}), consider
\begin{equation}
S := \sum_{m\ge 1}\frac{(-1)^m}{m}\sum_{k=1}^{m-1}\frac{(-1)^k}{k}
\prod_{j=1}^{k-1}\bigg(1+\frac{x}{j}\bigg),\label{pfmmn}
\end{equation}
the generating function for $\zeta(\ou,\ou,\us_n)$.
Since the inner sum of~(\ref{pfmmn}) is the generating
function for $\zeta(\ou,\us_n)$, we may write, in view of~(\ref{pfmn}),
\begin{equation}
S = \int_0^{-1} \frac{(1+u)^{-x}-1}{x(1-u)}\,du
  = \frac{1}{2}\int_0^1 \frac{1-u^{-x}}{x(1-u/2)}\,du
  = \sum_{k\ge 1} 2^{-k}\bigg(\frac{1}{k}-\frac{1}{k-x}\bigg),
\end{equation}
which is the right side of~(\ref{mmn}).

We factored the generating function~(\ref{all}) into linear factors and
then applied the infinite product representation for the Gamma function to
arrive at~(\ref{int}). In the same way, we arrived at~(\ref{nint})
from~(\ref{nall}).  The same procedure is done, in greater generality and
with more details provided, in~\cite{WW}, pp.~238--239.
Equations~(\ref{sin}) and~(\ref{cos}) arise from applying the reflection
formula for the Gamma function to~(\ref{int}) and~(\ref{nint})
respectively. Evaluations~(\ref{r2}) through~(\ref{r8}), and~(\ref{r10})
were derived from~(\ref{sin}) using the addition formulae to combine
products of sine functions into sums of trigonometric functions.  Likewise,
evaluations~(\ref{m2}) through~(\ref{m6}) were derived from~(\ref{cos}).

\raggedright

\end{document}